\documentclass[a4paper,11pt]{article}
\usepackage{authblk}
\usepackage[utf8]{inputenc}
\usepackage[english]{babel}
\usepackage{amsmath,amstext,amssymb}
\usepackage{xcolor}
\usepackage{bm}
\usepackage{graphicx}
\usepackage{textcomp}
\usepackage{url}
\usepackage{hyperref}
\usepackage{newlfont}
\usepackage[a4paper,left=2.5cm,right=2.5cm,bottom=4.0cm]{geometry}

\title{Emergent Gravity from an Augmented Variational Principle}
\author[a,b]{M. Tuveri}
\author[c,d]{L. Fatibene}
\author[c,d]{M. Ferraris}

\affil[a]{ -- Dipartimento di Fisica, Universit\`a di Cagliari\\Cittadella Universitaria, 09042 Monserrato, Italy}
\affil[b]{ -- INFN, Sezione di Cagliari}
\affil[c]{ -- Dipartimento di Matematica, University of Torino (Italy), via C. Alberto 10, 10123 Torino, Italy}
\affil[d]{ -- INFN Sezione  Torino -- Iniziativa Specifica QGSKY, via P. Giuria, 10125 Torino, Italy}

\begin{document}

\maketitle

email: \href{mailto:matteo.tuveri@ca.infn.it}{matteo.tuveri@ca.infn.it},
\href{mailto:lorenzo.fatibene@unito.it}{lorenzo.fatibene@unito.it},
\href{mailto:marco.ferraris@unito.it}{marco.ferraris@unito.it}

\begin{abstract}
A direct and non--trivial link between Padmanabhan's entropy used in emergent gravity and standard GR action is established.
To do that, Augmented Variational Principles (AVP) will be used.
We shall discuss how this link accounts for the details of the variation of Padmanabhan's action based on gravitational entropy.
It will also clarify the role of the background metric and its non-dynamical role.
\end{abstract}


\section{Introduction}
Quantum gravity is one of the open-issues that in the last decades has been studied extensively.
Gravity, which describes the behaviour of macroscopic objects as planets or galaxies,
also acts on quantum particles as it is manifest in black hole physics and its action needs to be described at least down to Planck scales. 
This suggests that we need a theory of quantum gravity to completely understand the physics of these astrophysical objects, 
even if the Planck scales (at which the back reaction of quantum particles on gravity is expected to become relevant) 
are currently way beyond our experimental capabilities. 

Accordingly, the physics of horizons (the three dimensional surfaces which surround black holes in spacetime) 
also is expected to be affected by quantum gravity. 
Thus solving the black hole puzzle may lead to a better understanding of the gravitational force at microscopic scales.

Although there are well developed proposals for quantum gravity, for example string theory or loop quantum gravity,
a completely satisfactory and accepted theory of quantum gravity is yet to come. 
Thus one is interested in different approaches to this topic. 

One of these is the thermodynamical approach, also called {\it emergent gravity}, mainly developed by Padmanabhan, see 
 \cite{PaddyThermo, PaddyQG, Chakraborty:2015aja, Padmanabhan:2016bha}, in the hope it could contribute to get information about 
quantum gravity bypassing the intricate mathematical difficulties of more general and fundamental approaches.

Emergent gravity is a classical approach and it offers a new viewpoint on the fundamental geometrical quantities one should 
use to develop a theory of quantum gravity \cite{PaddyQG}. 

The main idea of this approach is that horizons are totally observer dependent, so one can perceive them and
study their thermodynamical properties, even if there is no black hole; see \cite{Hawking}, \cite{Bekenstein1}, 
\cite{Bekenstein2}, \cite{Bekenstein3}.

In fact, we know that, for a Schwarzschild black hole, the surface $r=\frac{2Gm}{c^2}$ defines the horizon $\mathcal{H}$.
One can compute the so--called {\it surface gravity} $\kappa$ on $\mathcal{H}$ which corresponds to the gravitational acceleration at the horizon.
Therefore, in view of equivalence principle, starting from a flat spacetime, an observer sitting on spatial infinity 
can be accelerated along one direction, e.g. the $x-$direction, with
an acceleration $\kappa$ and, as a consequence, he will perceive a horizon. 
This kind of observers are called the \textit{Rindler observers} and they will 
attribute to this horizon every physical properties (the thermodynamic ones, too) as the observers in presence of a real horizon of a real black 
hole region.

This suggests to modify the Hamilton principle to obtain Einstein equations from quantities defining the horizons, namely the (lightlike) normal 
unit vector $n$ to the horizon, 
and related to thermodynamics. 

The Padmanabhan {\it ansatz} (see \cite{PaddyThermo}, \cite{PadVacuum}) 
is to use entropy as an action principle, and the normal $n$ to the horizon as a fundamental field, 
namely, in General Relativity
\begin{equation}
 S[n^a] = -\frac{c^4}{8\pi G}\int_Vd^4x\sqrt{-\bar{g}}\left[2P_{ab}{}^{cd}\nabla_cn^a\nabla_dn^b-\frac{8\pi G}{c^4}T_{ab}n^an^b -  
 \frac{8\pi G}{c^4}\lambda \cdot (n^a\bar{g}_{ab} n^b)\right] 
\label{PaddyS}
\end{equation}
where $\bar{g}$ is the determinant of the background metric $\bar{g}_{ab}$ and $V$ is a four dimensional volume of spacetime. Moreover 
the tensor $P_{ab}{}^{cd}= \frac{1}{2}\left(\delta_{a}^c\delta_b^d-\delta_a^d\delta_b^c\right)$ 
has all the symmetries of the Riemann tensor and it is a conserved tensor, viz., $\nabla_c P_{ab}{}^{cd}=0$, and
$T_{ab}$ is the (symmetric) energy-momentum tensor describing matter, which, of course, is also conserved,  viz., $\nabla_c T^{cd}=0$.
The function $\lambda$ has been added as a Lagrangian multiplier (to force $n$ to be lightlike) and the metric $\bar{g}_{ab}$ is here to be treated 
as a non-dynamical background.

Accordingly, this functional $S[n^a]$ is varied with respect to $n^a$ and $\lambda$ (to preserve the constraint for $n$ to be lightlike)
to obtain (see \cite{PaddyThermo})
\begin{equation}
 \begin{split}
  \delta S =&-2\frac{c^4}{8\pi G}\int_Vd^4x\sqrt{-\bar{g}}\left[2P_{ab}{}^{cd}\nabla_cn^a\nabla_d-
 \frac{8\pi G}{c^4}\left(T_{ab}+\lambda(x)\bar{g}_{ab}\right)n^a\right]\delta n^b\\
  =&-\frac{c^4}{4\pi G}\int_Vd^4x\sqrt{-\bar{g}}\left[-2P_{ab}{}^{cd}(\nabla_d\nabla_cn^a)-\frac{8\pi G}{c^4}(T_{ab}+
  \lambda(x)\bar{g}_{ab})n^a\right]\delta n^b\\
 &-\frac{c^4}{4\pi G}\int_{\partial V}d^3x\sqrt{\bar{h}}\left[2k_dP_{ab}{}^{cd}(\nabla_cn^a)\right]\delta n^b 
\label{variationS}
\end{split}
\end{equation}
where $k_d$ is a vector field normal to the boundary $\partial V$ and $\bar{h}$ is the determinant of the induced metric on $\partial V$. 
As usual, we require that the variations $\delta n^a$ should vanish on the boundary
so that the boundary part of the variation (\ref{variationS}) vanishes.

Accordingly, the condition for $S$ to be stationary for all the null vectors $n^a$ is 
\begin{equation}
 P_{ab}{}^{cd}\left(\nabla_c\nabla_d-\nabla_d\nabla_c\right)n^a=\frac{8\pi G}{c^4}\left(T_{ab}+\lambda(x)\bar{g}_{ab}\right)n^a 
\end{equation}
where we used the antisymmetry of $P_{ab}{}^{cd}$ in its upper two indices to write the first term. 
Thus, in General Relativity, the above expression becomes
\begin{equation}
 \left[R_{ab}-\frac{8\pi G}{c^4}\left(T_{ab}+\lambda(x)\bar{g}_{ab}\right)\right]n^a = 0
 \label{fieldentropy}
\end{equation} 
These are the field equations and they do not contain any derivative of the vector field $n^a$ and this peculiar 
feature arises because of symmetry requirements we imposed on the tensor $P_{ab}{}^{cd}$. 
Since this equation must hold for \textit{arbitrary} null vector fields $n^a$, then it becomes
\begin{equation}
 G_{ab} - \frac{8\pi G}{c^4}T_{ab}= \left[\frac{8\pi G}{c^4}\lambda(x)-\frac{1}{2}R\right]\bar{g}_{ab}\label{quasiEinstein}
\end{equation}
where we set $G_{ab}=R_{ab}-\frac{1}{2}R\bar{g}_{ab}$ for the {\it  Einstein tensor}.

By Bianchi identities and the conservation of $T_{ab}$, one has that $\partial_b\left[\frac{8\pi G}{c^4}\lambda(x)-\frac{1}{2}R\right]=0$, 
so the quantity $\Lambda=\frac{8\pi G}{c^4}\lambda(x)-\frac{1}{2}R$ is an integration constant.

The eq.(\ref{quasiEinstein}) then becomes
\begin{equation}
 G_{ab}= \frac{8\pi G}{c^4}\left(T_{ab}+ \frac{\Lambda c^4}{8\pi G}\bar{g}_{ab}\right)\label{EinstEq}.
\end{equation}
These are exactly the Einstein field equations in presence of matter and cosmological constant $\Lambda$ which, here, arises
as an integration constant which is partially determined by the value of the Lagrange multiplier we added to take into account 
that the null vectors $n^a$ must be null even after the variation of the functional; see \cite{PaddyThermo}, \cite{PadEntropy}. 

This thermodynamic approach to the problem of a definition of the black hole (spacetime) entropy allows to 
geometrically describe gravity as an \textit{emergent} phenomenon, since here the spacetime metric, 
emerges as the thermodynamic approximation of something of more fundamental than the metric itself, 
as a long wave approximation on the background $g$ of the spacetime infinitesimal degrees of freedom parametrised by $n$. 
For this reason this approach is called \textit{emergent gravity}; see \cite{PaddyThermo, PaddyQG, Chakraborty:2015aja, Padmanabhan:2016bha}.

Let us  notice that the Padmanabhan's ansatz (\ref{PaddyS})
gets a first mathematical rigorous definition by the work of 
Wald and Iyer \cite{EntropyWald} on conserved quantities in general covariant theories. 
In fact, they give there a definition of a gravitational 
entropy for dynamical black holes based on Noether charges, which, suitably manipulated, gives raise to the same Padmanabhan's 
gravitational functional. 

However, the Wald and Iyer results are based on a specific class of vectors (Killing vectors) and hypersurfaces on spacetime (bifurcate Killing horizons),
while Padmanabhan's quantities $n$ are unconstrained (except for being lightlike).
However, it has been shown that Wald and Iyer results can be obtained in a more general way without relying on special assumptions on horizons; 
see \cite{WaldNostro}, \cite{TaubNut} and references quoted therein.
These are also the starting point for a review of conservation laws in covariant theories which eventually lead to the so-called
{\it Augmented Variational Principles}, (AVP), \cite{Augmented}.

\medskip
We shall show how the null vector $n$ appearing in Padmanabhan's action (\ref{PaddyS})
is proportional to Wald's Killing vectors, and how Padmanabhan functional 
can be obtained from AVP by fixing the covariance gauge choosing a specific observer.

The material of this paper is organised as follows. 
In Section $2$ we shall review Augmented Variational Principles and show how one can recover in that context 
Padmanabhan's action (\ref{PaddyS}).
This also explains why Padmanabhan's action should be expected to produce Einstein equations (since the AVP does)
and in what sense the metric in Padmanabhan's action is to be considered as a non-dynamical object, 
while the normal vector $n$ is promoted to be dynamical.  

In Section $3$ we shall show the relation between the normal unit vectors $n$, the normal vector $\zeta$ (used in AVP) and Killing vectors $\xi$. 
Such a relation is not necessary (since Killing vectors do not play a special role 
in conservation laws for covariant theories; \cite{Symmetries}). However, it is relevant 
to render explicit the relations with Wald and Iyer results.

Finally, in the appendix we shall compute the Schwarzschild metric on the horizon for a Rindler observer since in the original 
computation \cite{PaddyThermo},
everything was evaluated in the near-horizon limit, 
thus leading to an approximated form of the metric which is not consistent with its exact form 
on the horizon. In fact the former turned out to be diagonal, while it is not.
However, the correction is irrelevant to the results in \cite{PaddyThermo}.

\bigskip

\section{The Augmented Variational Principle}

\textit{Augmented variational principles} (AVP) have been introduced in \cite{Augmented}, for a number of reasons, among which
they generalise the previous works on covariant conserved quantities for gauge natural field theories, 
they connect directly to the theory of variation of conserved quantities (see \cite{WaldNostro}), which in turn generalise the notion of geometric entropy
introduced in \cite{EntropyWald}. 
The interesting thing is that the generalisation is in the direction of making asymptotic behaviour (see \cite{BTZ1} and \cite{BTZ2}) 
and Killing horizons irrelevant (so that for example it applies to Taub Bolt solutions which has no horizon; see \cite{Taub1}, \cite{TaubNut}).
This is exactly the same direction suggested by the emergent gravity, and one can guess the AVP could be used
as a bridge between the two formulations and clarify the relations between them.

 AVP also improve the definition of energy associated with a black hole to solve in a natural way the problem of 
the so--called {\it Katz anomalous factor} (see \cite{Katz}) which, for example, in \cite{EntropyWald} is solved adding a 
suitable counter term in the Komar integral \cite{Komar} which leads to the correct result.

By using the AVP it is also possible to obtain field equations and the relative conserved 
quantities (for generic gauge covariant theories), i.e.~the amount of conserved quantity which has to be expended 
to drive a physical system from one $\bar{g}$ to another configuration $g$. 

In particular in General Relativity, setting $8\pi G=1$ and $c=1$, 
 this principle leads to an augmented gravitational Lagrangian (see \cite{Augmented}) 
\begin{equation}
  L^*=\sqrt{-g}R -\sqrt{-\bar{g}}\ \bar{R} - \nabla_k\left(\sqrt{-g}g^{ab}w_{ab}^k\right)
  \label{AugLag}
\end{equation}
where $R$ is the Ricci scalar defined by the metric $g$ and $\bar{R}$ is defined by $\bar{g}$. 
The total divergence depends on the tensor $w_{ab}^k= u_{ab}^k- \bar u_{ab}^k$ which 
is made up by the difference of the traceless connections  $u_{ab}^k = \Gamma_{ab}^k- \delta^k_{(a}\Gamma_{b)c}^c$ of the two metrics. 

The pure divergence is fundamental in the Lagrangian since, for example, it contributes to the definition 
of the Noether charge (viz., the superpotential) with an extra term which corrects the value of the black hole energy, correcting the anomalous factor
\cite{Augmented}. 
For computing relative conserved quantities, the two metrics need to induce the same metric at spatial infinity, 
though not necessarily being asymptotically flat.
The divergence can be considered as a covariant form for the counter term 
(and in fact it reduces to the Brown-York counter term depending on the extrinsic curvature 
on an ADM foliation; see \cite{BY}).

The action (\ref{AugLag}) is varied independently with respect to the metrics $g$ or $\bar{g}$ to get the field equations with respect to the metric $g$ 
or with respect to $\bar{g}$, respectively, which hence both obey Einstein equations.

\medskip

Now also forgetting about successes of the AVP, one could simply relate the action from (\ref{AugLag}) with the Padmanabhan's entropy (\ref{PaddyS}), 
in order to clarify their mutual relations.
For doing it we would like to write AVP, at least at the first order, as a function of 
some null vector defined in spacetime, thus leading to a functional which can be similar to Padmanabhan's entropy. 

Hence we should take the metric $g$ in AVP as the contribution of a background $\bar{g}$ and some null vector which parametrises the metric $g$
infinitesimally closed to the background $\bar g$.
Since in AVP the two metrics are decoupled, that explains why one can consider $\bar g$ fixed and vary only the null vector $n$, i.e.~$g$, 
and still obtain meaningful field equations. 
Accordingly, the reference metric $\bar g$ in the AVP will be eventually identified with the 
background metric appearing in the Padmanabhan's action (\ref{PaddyS}).\\

A Kerr-Schild metric (KS) \cite{Kerr-Schild, Kerr-Schild2, Kerr-Schild3, catalogo} 
is a good candidate for $g$ to show how the AVP depends on the difference $g-\bar g$ of two metrics. \\
In general, a metric $g$ is in {\it Kerr-Schild form} with respect to a background metric $\bar g$ 
(see \cite{Kerr-Schild, Kerr-Schild2, Kerr-Schild3, catalogo}), 
iff there exists a covector $\zeta_a$, 
\begin{equation}
 g=\bar{g} +\zeta\otimes \zeta 
 \label{KSgeneral}
\end{equation}
where the covector $\zeta_a$ is such that the vector $\bar{\zeta}^a=\bar{g}^{ab}\zeta_b$ is lightlike with 
respect to $\bar g$ (that is $\zeta_a\bar \zeta^a=0$). One can easily check that the inverse 
metric is $g^{ab}=\bar{g}^{ab}-\bar \zeta^a\bar \zeta^b$. 
Therefore the vector $\zeta^a:=g^{ab}\zeta_b$ is
\begin{equation}
 {\zeta}^a:=g^{ab}\zeta_b=\bar{g}^{ab}\zeta_b-\bar \zeta^a\bar \zeta^b\zeta_b=\bar{\zeta}^a
\end{equation}


This vector is rather special since it is a light-like
vector for the metric $g$ and also for the metric $\bar{g}$, since $\zeta^a=\bar{\zeta}^a$ and it is 
globally defined everywhere in spacetime.\\

KS metrics seem to be quite relevant in the context of classical and quantum gravity at semi-classical level. 
From a classical point of view, when the background metric $\bar{g}$ is taken to be the Minkowski spacetime, 
KS type vacuum solutions can be found by solving the linearized vacuum Einstein
equation \cite{Xanthopoulos:1978, Gergely:2002hm}. Further, it has been shown 
\cite{Gibbons:2004uw, Burinskii:2012} that, KS metrics
include many important curved space-times of General Relativity, such as charged and rotating black holes and stars (with and without horizon), 
de Sitter and Anti de Sitter space-times and their rotating analogues, and so on. The 
KS ansatz \eqref{KSgeneral} appears to be valid not only in General Relativity, 
but also in more general higher derivatives theory like Lovelock gravity
\cite{Ett:2011fy}.\\
From a quantum gravity point of view instead, in the context of black holes one finds that, for example, KS metrics 
also give new insights on the structure of quantum horizon and mechanism of black hole evaporation \cite{Burinskii:Milan, Burinskii:2009wx} and 
also that it is possible to find logarithmic correctionts to their entropy \cite{El-Menoufi:2015cqw} in agreement to what is commonly believed 
about black holes entropy \cite{Jacobson:1994, Jacobson:1993}.\\

For example, we can write the Schwarzschild metric $g$ (of mass $m$) in the Kerr-Schild form 
with respect to a (different) Schwarzschild background metric $\bar g$ (of mass $\bar m$). 
By understanding  angular directions, let us consider a transformation $dt=dt'-f(r)dr$ on the metric $g$, 
and recast it as
\begin{equation}
 g=\bar{g}+\frac{r-2\bar{m}}{r}dt^{'2}-\frac{rdr^2}{r-2\bar{m}}-\frac{r-2m}{r}\left(dt'-f(r)dr\right)^2+\frac{r}{r-2m}dr^2
\label{SchwSchwKS}
\end{equation}
where 
\begin{equation}
 \bar{g}=-\left(1-\frac{2\bar{m}}{r}\right)dt^{'2}+\frac{dr^2}{1-\frac{2\bar{m}}{r}}.
\end{equation}
Here $g$ and $\bar{g}$ only differ for their mass, which are $m=\bar{m}(1+\delta m)$ and $\bar{m}$, 
respectively, and $\delta m$ is an infinitesimal parameter. 
Since we wish $g$ to be in Kerr-Schild form
 we search for a function $f(r)$ such that we can write $g-\bar g$ as the square of a (null) covector $\zeta$. 
Hence this function is $f(r)=\pm\frac{2r(m-\bar{m})}{(r-2m)(r-2\bar{m})}$ and the difference $g-\bar{g}$ becomes
\begin{equation}
 g-\bar{g}=\frac{2(m-\bar{m})}{r}\left(dt^{'}\pm\frac{r}{r-2\bar{m}}dr\right)^2.
\end{equation}
By comparing with (\ref{SchwSchwKS}) we have two possible (future directed) covectors $\zeta_{\pm}$, which expression is
\begin{equation}
\begin{split}
 \zeta_{\pm}&=\sqrt{\frac{2(m-\bar{m})}{r}}\left(dt^{'2}\pm\frac{r}{r-2\bar{m}}dr\right)
 =\sqrt{\frac{2(m-\bar{m})}{r}}\left(dt^{2}\pm\frac{r}{r-2m}dr\right)\label{Schwzetapm}
 \end{split}
\end{equation}
where in the last line we used the expression of $f(r)$ to switch from the coordinates $(t',r)$ back to $(t,r)$.
So, the Kerr-Schild metric becomes
\begin{equation}
 g=\bar{g}+\zeta_{\pm}\otimes\zeta_{\pm}.
\label{KSgbar}
\end{equation}
Accordingly, we can expand a Schwarzschild metric $g$ as a Kerr-Schild metric with respect to another 
Schwarzschild metric $\bar g$ and the null vector $\zeta_\pm$ parametrises infinitesimal displacements along the family of 
Schwarzschild solutions.

Thus, our aim is to write the 
AVP for a Kerr-Schild metric and, eventually, compare the result to Padmanabhan's action (\ref{PaddyS}).
In anycase for the reasons written 
above, we believe that our procedure is not only valid for a Schwarzschild metric, for which we have explicit shown it, but it 
can be extended to (and it should be valid for) a wide class of metrics, thus making more general our results.
\medskip

Now we can check what is the general expression of the AVP (in standard GR) for a Kerr-Schild metric (\ref{KSgeneral}). 
The Ricci scalar  can be written as (see \cite{catalogo})
\begin{equation}
 R=\bar{R}+\nabla_a\left(2\zeta^a\nabla_b\zeta^b\right)-2P_{ab}{}^{cd}\nabla_c\zeta^a\nabla_d\zeta^b \label{RicciKS}
\end{equation}
and since the choice of the background metric $\bar g$ is arbitrary, we can set $\bar{g}$ as a vacuum solution (e.g. Minkowski metric), 
so that one has $\bar{R}=0$,
and $g$ should asymptotically agree with $\bar{g}$.
Hence the augmented gravitational 
action becomes, at least at first order, a function of the null vectors $\zeta^a$.

Substituting this scalar curvature
in the AVP (\ref{AugLag}) we can define an 
augmented gravitational action by integrating over a spacetime region $V$ and one easily gets
\begin{equation}
 A_{grav}^{*}=\int_Vd^4x\left[\left(-2\sqrt{-g}P_{ab}{}^{cd}\nabla_c\zeta^a\nabla_d\zeta^b\right)
 +\nabla_c\left(2\sqrt{-g} P_{ab}{}^{cd}\zeta^a\nabla_d\zeta^b\right)\right].
 \end{equation}
We notice that the first term on the right-hand side of the 
action is a bulk term, whereas the second one is a total divergence that, as  discussed above, 
does not contribute to the field equations of the theory.

Moreover in order to define a total action, we should also give an explicit form for the matter action which is not trivial, so our proposal 
is based on a purely thermodynamic reasoning. Let us remember that the Lagrangian density 
is a mathematical quantity which exhibits all the features of an energy density, since it has its dimensions.

Let us consider a perfect fluid described by an energy-momentum tensor $T_{ab}$.
The integral curves of $\zeta^a$ must be null geodesics of $g$, thus  $\zeta^a\nabla_a\zeta^b=0$ 
(which also satisfies the null energy conditions).
Then the energy density has the following expression:
\begin{equation}
 \rho=T_{ab}\zeta^a\zeta^b
\end{equation}
where the vector $\zeta^a$ is the future-directed spatial hypersurface normal vector.
This kind of hypersurface can be, for example, the horizon of a black hole region.

This allows us to write the 
contribution of the matter action as
\begin{equation}
 A_{mat}^{*}=\int_Vd^4x \sqrt{-g} T_{ab}\zeta^a\zeta^b
\end{equation}
Let us also notice that, if the time integration is associated to the inverse of the temperature
of the region over which we integrate (for example by considering the Euclidean extension of spacetime, see \cite{PaddyTemp}),
then this action can be interpreted as the matter contribution to the entropy in this region. Similarly, the same holds for the gravitational AVP, 
which could be thought as the gravitational contribution to the entropy in $V$. Thus we are allowed to write the 
total entropy over a certain region of spacetime as
\begin{equation}
 \begin{split}
A_{tot}[\zeta^a]=&-\frac{c^4}{8\pi G}\int_Vd^4x\sqrt{-g}\left[2P_{ab}^{cd}\nabla_c\zeta^a\nabla_d\zeta^b
 -\frac{8\pi G}{c^4}T_{ab}\zeta^a\zeta^b\right]\\
 &+\frac{c^4}{8\pi G}\int_Vd^4x\nabla_c\left(2\sqrt{-g}\epsilon P_{ab}^{cd}\zeta^a\nabla_d\zeta^b\right)\label{OURS}
 \end{split}
 \end{equation}
where we restored the constant $\frac{c^4}{8\pi G}$ to make a simpler comparison between this action 
and the Padmanabhan's functional in eq.(\ref{PaddyS}).
The augmented action is a function of the null vectors $\zeta^a$ and since it has the same form of the Padmanabhan's entropy functional, 
the field equations follow as in the previous Section.\\
We note that the choice to set $\bar{g}$ as a vacuum solution is just to make evident the comparison between Padmanabhan's 
ansatz \eqref{PaddyS} and our augmented action. 
In fact, with a generic curved spacetime, the (background) Ricci scalar, $\bar{R}$, contributes to the total action \eqref{OURS}, 
but it does not to the field equations, since it does not depend on the null vectors $\zeta$. 

It is clear that changing the null vector $\zeta$ correponds to change the metric $g$, at least at the first order.
We still have to prove that if the change of $\zeta$ preserves null vectors, then the new metric is still a solution.
Once this has been shown, then there is a complete parallel between the variation of the AVP with respect to $g$ and
the awkward way the Padmanabhan's action is varied.

\section{The relation between $n^a$ and $\zeta^a$}

In order to compare the augmented action in eq.(\ref{OURS}) with the entropy functional in eq.(\ref{PaddyS}) 
we need to explicitly see if the globally defined null vectors $\zeta^a$ are 
in some way related with $n^a$ (which are null vectors only on $\mathcal{H}$). At first we will show the form of the null vectors $n^a$, 
then the one of ${\zeta}^a_{\pm}$.\\ 
Following the Padmanabhan algorithm given in \cite{PaddyThermo}, we start by considering
the Rindler form for static metrics which can be written as
\begin{equation}
 ds^2 = -N^2dt^2 +\gamma_{\alpha\beta}dx^{\alpha}dx^{\beta}.\label{generalRindler}
\end{equation}
where the function $N$ and the spatial metric $\gamma_{\alpha\beta}$ just depend on $x^\alpha$ not on time.

Then we can define the comoving observers as the ones moving with trajectories $x = constant$ (and $\theta = \phi = const$) which have four velocity 
$u_a = -N\delta_a^0$ and four acceleration $a^i = u^j\nabla_ju^i = (0,\vec{a})$, which has the purely spatial components 
$a_{\alpha} = (\partial_{\alpha}N)/N$. In general, the unit normal $n_{\alpha}$ to the $N=constant$ surface is given by 
\begin{equation}
n_{\alpha}=\partial_{\alpha}N(g^{\mu\nu}\partial_{\mu}N\partial_{\nu}N)^{-1/2}=a_{\alpha}(a_{\beta}a^{\beta})^{-1/2}\label{n} 
\end{equation}
This vector has a useful expression when related to another important vector that static metrics admit: in fact 
they admit a timelike Killing vector 
field $\xi^a$ which has components ${\xi}=(1,0,0,0)=\partial_t$
in Rindler coordinates. 
The norm of this vector vanishes on the horizon, which - in turn - acts as a bifurcation surface. The vector ${\xi}$ can be helpful to give
a definition of surface gravity $\kappa$ through its acceleration
\begin{equation}
 \xi^a\nabla_a\xi^b = \kappa\xi^b.
\end{equation}
In the freely falling frame (see eq.(\ref{PADDYFLAT}) in Appendix \ref{App:AppendixA} below), which is also a
locally inertial frame with Cartesian coordinates ($T,X$), both
the normal to the $N=constant$ surfaces and the velocity of the comoving observers become parallel to the Killing vector $\xi^a$, 
when they are evaluated on the horizon, $\mathcal{H}$.
In fact, let $I$ denote the inertial frame, whose bases vectors are $(\partial_T, \partial_X)$ and $R$ denote the Rindler frame, 
 whose bases vectors are $(\partial_t,\partial_r)$; see the appendix \ref{App:AppendixA}. Then \cite{PaddyThermo}:
\begin{itemize}
 \item The Killing vector field $\xi^a$ has the components $\xi = \partial_t$ in the Rindler frame or, in the inertial frame, 
 \begin{equation}
  \xi = \kappa(X\partial_T+ T\partial_X)
 \end{equation}
and its norm vanishes on the horizon which is defined by $N^2=0$, since $\xi_a\xi^a = -N^2$. 


\item The normal $n^a$ to the $N=constant$ surfaces is ill-defined on the horizon, thus we consider the redshifted normal which has 
the following limit on $\mathcal{H}$
\begin{equation}
  Nn^a|_{\mathcal{H}} \equiv \xi^a|_{\mathcal{H}}=\frac{X}{4m}(\partial_T+\partial_X) \label{Nn}
 \end{equation}
and it is also a null vector on $\mathcal{H}$, since the Killing vector is. 
\end{itemize}

These vectors play a fundamental role in the definition of the Padmanabhan's entropy functional, since they represent the null vectors 
through which it is 
possible to parametrize the horizon internal degrees of freedom.\smallskip

Now we will give the expression of the globally defined null vectors ${\zeta^a_\pm}$.\\ 
Given the Schwarzschild metric, choosing the background $\bar{g}$ as a Minkowski metric, $\eta$, the Kerr-Schild form for this metric is
\begin{equation}
 g=\eta +\zeta_{\pm}\otimes\zeta_{\pm} \label{KSSchwonMink}
\end{equation}
where $\eta=-dt'^2+dr^2$ and $\zeta^a_\pm$ are given \begin{equation}
 {\zeta_\pm}=\sqrt{\frac{2m}{r}}\left(-\frac{r}{r-2m}\partial_t\pm\partial_r\right).
\end{equation}
These vectors are defined everywhere in spacetime and they are also null vectors (both for $g$ and $\eta$), in fact their norm is
\begin{equation}
 g({\zeta}_{\pm},{\zeta}_{\pm})=-\frac{2m}{r}\frac{r-2m}{r}\frac{r^2}{(r-2m)^2}+\frac{2m}{r}\frac{r}{r-2m}=0.
\end{equation}

Let us start by considering the vector ${\zeta}_{-}$, which expression is
\begin{equation}
 \zeta_{-}=-\sqrt{\frac{2m}{r}}\left(\frac{r}{r-2m}\partial_t+\partial_r\right).
\end{equation}
In order to write its expression on the horizon, first we move in the Rindler frame: 
using the space coordinate transformation (see Appendix \ref{App:AppendixA} to the correct definition 
of the Schwarzschild metric in the Rindler frame and in the freely falling frame, too)
$r=2m(x^2+1)$, which leads to the partial space derivative transformation $\partial_r = \frac{1}{4mx}\partial_x$, and the time coordinate is left 
unchanged, this null vector becomes
\begin{equation}
 {\zeta}_{-} = -\sqrt{\frac{1}{x^2+1}}\left(\frac{x^2+1}{x^2}\partial_t+\frac{1}{4mx}\partial_x\right)
\end{equation}
Now, we want to compare the Rindler vector bases with the freely falling ones, so using the eq.(\ref{OURCOORDTRANSF}) of Appendix \ref{App:AppendixA} 
and the chain rule we get
\begin{equation}
\partial_x=\frac{1}{x}\left(T\partial_T+X\partial_X\right),\qquad \partial_t = \frac{1}{4m}\left(X\partial_T+T\partial_X\right) \label{partialtx>TX} 
\end{equation}
then, substituting them into the expression of ${\zeta}_{-}$, also with the relation $x^2=\frac{X^2-T^2}{16m^2}$, we have
\begin{equation}
 {\zeta}_{-}
 =-\sqrt{\frac{16m^2}{X^2-T^2+16m^2}}\left[\left(\frac{X}{4m}+\frac{4m}{X-T}\right)\partial_T+\left(\frac{T}{4m}+\frac{4m}{X-T}\right)\partial_X\right].
\end{equation}
This vector has not a finite limit on the horizon: in fact, on the part of the horizon we are interested in, when $|X|<T$, 
that is where the horizon is defined by $T= X$, the vector ${\zeta}_{-}$ becomes
\begin{equation}
 {\zeta}_{-}=
 -\frac{4m}{X-T}(\partial_T+\partial_X)
\end{equation}
and it diverges. 
However, on the horizon, it is possible to rescale this vector so that its relation with ${n}|_{\mathcal{H}}$ clearly arises and in fact it 
can be rewritten as
\begin{equation}
 {\zeta}_{-}|_{\mathcal{H}}\to \nu_{-}|_{\mathcal{H}}=-\frac{X^2-T^2}{32m^2}\zeta_{-}=Nn\big|_{\mathcal{H}}=\xi\big|_{\mathcal{H}}.\label{zetan}
\end{equation}
Thus the suitably rescaled Kerr-Schild null vector ${\nu}_{-}\big|_{\mathcal{H}}$ has a relationship with the Killing vector field of the 
Schwarzschild metric, so they are vectors which lie on the same branch of $\mathcal{H}$, since they are proportional and (anti-)parallel.\\
Finally, it can be shown that the vectors ${\zeta}_{+}$ are not related to the null vectors $Nn|_{\mathcal{H}}$ since they lie on a different branch of 
the light-cone, that is where $|X|>T$.\\
We notice that when we defined the augmented action in eq.(\ref{OURS}) through the null vectors ${\zeta}$ we refer to them as the null ones, 
${\zeta}_{-}$, shown above.

\section{Conclusions}

The Augmented Variational Principle \cite{Augmented} is a very useful tool in general
covariant theories as standard GR for solving some issues regarding
the definition of a entropy functional and it gives a very general definition
of this quantity.\\
In fact, given a Kerr-Schild metric of the form (\ref{KSgeneral}), 
we have shown that it is possible to establish a not trivial relation between the AVP in eq.(\ref{OURS})
and the Padmanabhan's entropy  functional in eq.(\ref{PaddyS}). 
Since the latter has the same form of the former, now it is possible to give a rigorous and general 
mathematical foundation to the 
Padmanabhan's ansatz for entropy in GR:
\begin{enumerate}
 \item The physical intuition based on the crucial role of null vectors 
 to define an entropy functional has now a mathematical foundation in the choice of a Kerr-Schild form of the metric to 
 describe black holes. Given a Schwarzschild metric written in a Kerr-Schild form with the background metric set as a vacuum solution, 
 then, at least at the first order, the augmented action can be written as a function of some globally 
 defined null vectors, ${\zeta}$. 
 \item 
 It was not clear why the variation of an entropy functional should lead to certain field equations for a given theory. 
 On the other hand it is 
 clear that the variation of an action functional leads to some field equation. However, since the Padmanabhan entropy has the same 
 form of the augmented action, the variation of these functionals 
 leads to the same field equations shown in eq.(\ref{EinstEq}), which - in turn - have the form of the Einstein's one 
 in presence of matter and a cosmological 
 constant which, here, arises as an integration constant which takes part of its value from adding of Lagrange multiplier to take into 
 account that the 
 null vectors $\zeta$ stay null even after the variation of the action (we give a geometrical explanation and justification of this fact 
 in the Appendix \ref{App:AppendixA}). 
 \item We also notice that there are some differences between the functional in eq.(\ref{OURS}) and the one in eq.(\ref{PaddyS}): 
 in the augmented gravitational action it also appears a total divergence term which leads to a surface term but, in order to get a well defined
 variational principle, we can ask that the variation of the vectors $\delta\zeta^a$ vanishes on the boundary. This request makes sense, since one of 
 the properties of the globally defined Kerr-Schild null vector is that they have the same expression for the metric $g$ and $\bar{g}$, thus 
 $\delta\zeta^a=\zeta^a-\bar{\zeta}^a=0$ with $\zeta^a$ and $\bar{\zeta}^a$ null vectors both for $g$ and $\bar{g}$. 
 Thus the bulk terms have the same form. In AVP, the surface term arising from a integration by parts of the bulk term has
 the same sign of the one in Padmanabhan algorithm, while our total divergence term is defined as positive.\\
 The other difference is that the augmented action is a function of the null vectors $\zeta^a$, whereas the Padmanabhan's entropy is a function of $n^a$. 
 However, the relation between these two vectors is given in eq.(\ref{zetan}) and we have shown that given a point on the horizon, 
 they lie on the same branch of the light-cone at that point. Moreover, when we minimize the AVP to find the field equations of the theory, 
 the rescaling factor which relates these two vectors factorizes itself, leading to the same Padmanabhan's field equations. 
 We expect a result of this type, since the field equations must hold for arbitrary null vectors defined on $\mathcal{H}$.
 \item Finally, when evaluated on-shell, the bulk contribution of the AVP must vanish, thus leading to a purely surface term which is
\begin{equation}
 A_{grav}^*=\int_{\partial V}d^3xk_c\left(2\sqrt{h}P_{ab}{}^{cd}\zeta^a\nabla_d\zeta^b\right).
\end{equation}
As we have shown in the previous section, it is possible to relate the null vector $\zeta$ to the Killing vectors $\xi$ by using a certain 
function of the coordinates $f(x)= -\frac{X^2- T^2}{32m^2}$
given by (33). In this way, one can associate an entropy to the surface action: in fact, the latter becomes 
\begin{equation}
\begin{split}
 A_{grav}^*&=\int_{\partial V}d^3xk_c\left[2\sqrt{h}P_{ab}{}^{cd}f(x)^{-1}\xi^a\nabla_d\left(f(x)^{-1}\xi^b\right)\right]\\
&=\int_{\partial V}d^3xk_c\left[2\sqrt{h}P_{ab}{}^{cd}f(x)^{-1}\xi^a\left(-\frac{\nabla_d f(x)}{f^2(x)}\xi^b+f(x)^{-1}\nabla_d\xi^b\right)\right]
 \end{split}
\end{equation}
Using the definition of the tensor $P_{ab}{}^{cd}$, the first term into the parentheses vanishes and the previous expression becomes
\begin{equation}
A_{grav}^*=\int_{\partial V}d^3x\left[2f(x)^{-2}\sqrt{h}\left(k_a\xi^a\nabla_b\xi^b-k_b\xi^d\nabla_d\xi^b\right)\right].
\end{equation}
The Killing vector $\xi$ satisfies the condition 
$\xi^d\nabla_d\xi^b=\kappa\xi^b$ which, in turn, defines the surface gravity on the surface $\partial V$.
Since $\xi^a$ is a Killing vector satisfing the Killing equation $2\nabla_{(a}\xi_{b)}=0$, the first term into the parentheses vanishes and the surface 
action becomes
\begin{equation}
 A_{grav}^*=-\int_0^{2\pi/\kappa}dt\int_{\partial V}d^2x\sqrt{\sigma}f(x)^{-2}k_b\xi^d\nabla_d\xi^b
\end{equation}
where ${2\pi/\kappa}$ is a quantity related to the (inverse of the) temperature; see \cite{PaddyTemp}, 
$\sigma$ is the determinant of the induced metric on $\partial V$ and 
$k_c$ is a null covector satisfying the relation $k_b\xi^b=-f(x)^2$. The action becomes
\begin{equation}
 A_{grav}^*=\frac{1}{4}A_{\mathcal{H}}
\end{equation}
where we have restored the constant $8\pi$, with $G=1$. This is the well known formula (\cite{Hawking}) 
for the entropy of a Schwarzschild black hole, also obtained by Padmanabhan in \cite{PaddyThermo}.
\end{enumerate}
This shows that the AVP is a powerful mathematical tool by which one can describe the dynamics of spacetime when 
searching for the relations between two metrics, one of them chosen as a background, the other used to describe spacetime dynamics. 
In General Relativity, with a Schwarzschild 
spacetime involving black holes, the Kerr-Schild metric is perfect to express in a natural way this kind of relation (see also \cite{Bini:2014nga, 
Gurses:2016moi} for 
recent developments in the study of these metrics). 
This metric is a function of globally defined null vectors and as a consequence, 
the augmented action becomes a function 
of these vectors. Thus the metric does not play any role as a dynamical field (at least at the first order), 
varying this action, we get the Einstein field equations and when they are satisfied, the boundary terms of the action lead to the entropy 
of a black hole  
(if the integration on the time coordinate is in same way related to the Hawking temperature of the black hole \cite{PaddyThermo}).\\
Let us finally remark that the relation between Padmanabhan's entropy action and AVP is not trivial. While in fact, 
in Padmanabhan's argument the Rindler form of the metric, the horizon, the null-vector on the horizon play a fundamental role, in AVP 
they play almost no role at all, still leading though essentially to the same results.

Let us note that in the last years, Padmanabhan has provided a formulation of the standard General Relativity action 
\cite{Kothawala:2014fva, Kothawala:2014tya} based on some 
geometrical quantities called ``qmetrics'' \cite{Kothawala:2013maa} which, suitably manipulated, 
leads to a functional which has the same form of 
the entropy ansatz given in \cite{PaddyThermo}. This is a very interesting fact, since this new way to rewrite General Relativity 
starts from non-local geometrical quantities defined in a \textit{mesoscopic} regime at the Planck scales (defined by the Planck lenght $L_P$), 
where the notions of differential 
geometry and quantum mechanics are still valid \cite{Kothawala:2014fva, Kothawala:2014tya}. 
However, in the classical regime where the quantum effects are negligible ($L_P\to0$), 
one expects the non local action to reduce to the Hilbert-Einstein one which, on the contrary, is based on local quantities (the metric tensor). 
To do so, one has to define a suitable limit procedure to move from the ``mesoscopic'' action to the classical one 
\cite{Kothawala:2014fva, Kothawala:2014tya, Stargen:2015hwa}. Then the standard variational principle naturally applies, thus leading to 
the Einstein field equations as in \cite{PaddyThermo}.\\
On the other hand, AVP is a non local variational principle, since it relates the dynamics of two different local metrics ($g$ and $\bar{g}$) and, 
together with the Kerr-Schild ansatz, allows us to find that the Hilbert-Einstein action can be written as the Padmanabhan's 
entropy ansatz; in fact, the augmented action is a function of some null vectors 
(well-defined everywhere in spacetime) which can encode the non locality features of a quantum spacetime in a mesoscopic regime 
as considered in \cite{Kothawala:2013maa, Kothawala:2014fva, Kothawala:2014tya, Stargen:2015hwa}.
Further investigations can shed light on the relationship between these two approaches.

Of course, Padmanabhan's argument may well have important physical motivations, however, in view of the mathematical equivalence with AVP,
one has to accept that, from a mathematical viewpoint, they are unessential 
and one can provide a simpler account of Padmanabhan's results.

As a consequence of our results, 
if the emergent gravity perspective to study the physics of internal degrees of freedom of spacetime has some sense, 
the AVP could give a better and clearer mathematical foundation to this physical intuition. 

\appendix
\section{Appendix: The Horizon Limit of the Schwarzschild Metric}

\label{App:AppendixA}
In this appendix we set the constants $8\pi G=1$ and also $c=1$.

Let us remember the starting point of Padmanabhan's approach: the concept of horizon is observer dependent, thus 
if, on one hand, the Schwarzschild metric exhibits a pathological
behavior at $r=2m$, which defines the location of the horizon, on the other hand 
an observer can perceive this kind of surface starting from the flat 
spacetime, where the metric is
\begin{equation}
 g=-dT^2+dX^2+dL_{\perp}^2\label{PADDYFLAT}
\end{equation}
where $(T,X)$ are Cartesian coordinates which are everywhere well-defined, then being accelerated with acceleration $\kappa$ along the $x-$direction. 
Using the coordinate transformation
\begin{equation}
 \kappa X=\pm\sqrt{2\kappa|l|}\cosh(\kappa t) \qquad \kappa T = \pm\sqrt{2\kappa|l|}\sinh(\kappa t) \label{coordtransf}
\end{equation}
the metric becomes
\begin{equation}
 g = -2\kappa l dt^2 + \frac{dl^2}{2\kappa l} + dL_{\perp}^2 
\end{equation}
where $l=0$ represents the location of the horizon. Setting $l=\frac{1}{2}\kappa x^2$, the horizon is at $x=0$ and the metric now becomes
\begin{equation}
 g=-\kappa^2x^2dt^2+dx^2+dL_{\perp}^2.\label{RINDLERPADDY}
\end{equation}
In particular, we are interested in studying the normal vectors to the horizon.

Let us notice that 
the Schwarzschild coordinates are not defined on the horizon $r=2m$, since some of the coefficients of the metric
are ill-defined on that surface, thus to describe the physics involving horizon one should switch to other coordinate frames. 
In the freely falling frame, the coordinates $(T,X)$ are well-defined 
around the horizon, thus one should switch from the Schwarzschild coordinates to the Rindler ones, then to the freely falling ones, i.e.~to 
$(T,X)$ coordinates.
In fact, 
the Rindler form of the metric in 
eq.(\ref{RINDLERPADDY}) represents also a general form for static metrics as the Schwarzschild one to be written \cite{PaddyThermo}, 
thus we expect the latter to have 
the form in eq.(\ref{RINDLERPADDY}) in Rindler coordinates, and that in eq.(\ref{PADDYFLAT}) in the freely falling one, when we evaluate it  on the horizon 
(at $X=\pm T$).

Understanding the angular coordinates, the Schwarzschild metric reads as
\begin{equation}
 g=-\left(1-\frac{2m}{r}\right)dt^2+\frac{dr^2}{1-\frac{2m}{r}},
\end{equation}
Then using the coordinate transformations $r=2m(x^2+1)$ and $dr=4mxdx$, with the time coordinate resting unchanged, the Rindler expression 
for this metric is
 \begin{equation}
 g = -\frac{x^2}{x^2+1}dt^2 + 16m^2(x^2+1)dx^2.\label{OURRINDLERMETRIC}
\end{equation} 
The Rindler coordinates are related to the Cartesian ones by these relations
\begin{equation}
 X=4mx\cosh\left(\frac{t}{4m}\right),\qquad T=4mx\sinh\left(\frac{t}{4m}\right)\label{OURCOORDTRANSF}
\end{equation}
and the Schwarzschild metric becomes 
\begin{equation}
\begin{split}
 g =&\frac{1}{16}\frac{T^4-T^2X^2-32T^2m^2+256m^4}{m^2\left(T^2-X^2+16m^2\right)}dT^2
 -\frac{1}{8}\frac{TX\left(T^2-X^2-32m^2\right)}{m^2\left(T^2-X^2-16m^2\right)}dXdT + \\
 &+\frac{1}{16}\frac{T^2X^2-X^4-32X^2m^2-256m^4}{m^2\left(T^2-X^2+16m^2\right)}dX^2.\label{OURFREELYPADDY}
\end{split}
 \end{equation}
In this frame the horizon is located at $T=\epsilon X$, where $\epsilon=\pm 1$. On the horizon, the metric is well defined and 
it has the following expression
\begin{equation}
 g|_{\mathcal{H}}=\frac{X^2-8m^2}{8m^2}dT^2-\frac{\epsilon X^2}{4m^2}dXdT +\frac{X^2+8m^2}{8m^2}dX^2.\label{OURFREELYPADDYH}
\end{equation}
The comparison between our results and those in eq.(\ref{RINDLERPADDY}) and eq.(\ref{PADDYFLAT}) is immediate:
\begin{itemize}
 \item The Rindler metric in eq.(\ref{RINDLERPADDY}) can be obtained from the Schwarzschild at first setting $r-2m=\frac{\tilde{x}^2}{8m}$ and 
 $dr=\frac{\tilde{x}d\tilde{x}}{4m}$. Then, in order to write this metric near the horizon, we expand in series the denominator of this equation at the point
 $x=0$. At order $x^4$, the Rindler metric becomes
 \begin{equation}
  g=-\kappa^2\tilde{x}^2dt^2+d\tilde{x}^2 +o(\tilde{x}^4)
 \end{equation}
 where the $\tilde{x}$ coordinate and the $x$ one are related by $\tilde{x}=4mx$. 
 This metric is different from the one in eq.(\ref{OURRINDLERMETRIC}), which we have obtained without any kind of approximation. In fact, the series 
 expansion at the point $\tilde{x}=0$ ($x=0$) is wrong, since the Rindler coordinates are not defined at this point. The eq.(\ref{RINDLERPADDY}) 
 cannot be used to represent the Schwarzschild metric in Rindler coordinates ($t,x$), 
 while our metric is well-defined in this frame (but not at $x=0$), so it can be used for this
 purpose.   
 \item Due to this bad approximations, in the freely falling frame, the difference between the metric in eq.(\ref{OURFREELYPADDY}) 
 and the one in eq.(\ref{PADDYFLAT}) becomes more evident. In fact, when we evaluate our Schwarzschild metric in the freely falling frame, its limit
 on the horizon is well-defined and shows the presence of a term in $dXdT$ which the eq.(\ref{PADDYFLAT}) does not show up; moreover, 
 the terms proportional
 to $dX^2$ and $dT^2$ are not linear. Nevertheless, when one wants to expand the denominator of the Rindler metric at the point $x=0$, 
 it should consider the metric in eq.(\ref{OURRINDLERMETRIC}) and expand at order $x^8$. In this case the Rindler metric becomes
 \begin{equation}
  g \simeq -x^2(1-x^2-x^4-x^6)dt^2+16m^2(x^2+1)dx^2+o(x^{10}).
 \end{equation}
 Then, using the coordinate transformation in eq.(\ref{OURCOORDTRANSF}), we can write an approximated Rindler metric in the freely falling frame, 
 getting, at order $o((X^2-T^2)^3)$, 
\begin{equation}
\begin{split}
 g = &-\frac{1}{16m^2}\Bigg\{16m^2-(X^2+T^2)-X^2\left[\frac{X^2-T^2}{16m^2}-\left(\frac{X^2-T^2}{16m^2}\right)^2\right]\Bigg\}dT^2\\
&+2\frac{XT}{16m^2}\left[-2-\frac{X^2-T^2}{16m^2}-\left(\frac{X^2-T^2}{16m^2}\right)^2\right]dXdT\\
 &-\frac{1}{16m^2}\Bigg\{-16m^2-(X^2+T^2)-T^2\left[\frac{X^2-T^2}{16m^2}-\left(\frac{X^2-T^2}{16m^2}\right)^2\right]\Bigg\}dX^2
\end{split}
\end{equation}
where the factor $2$ in the mixed term is due to the fact that the product $dXdT$ is symmetric. 
When evaluated on $\mathcal{H}$, $X=\pm T$, the sixth and eighth order terms vanish since they are of order grater than one in $(X^2-T^2)$,
thus leading to
\begin{equation}
 g|_{\mathcal{H}} = \left(-1+\frac{X^2}{8m^2}\right)dT^2 -\frac{X^2}{4m^2}dXdT + \left(1+\frac{X^2}{8m^2}\right)dX^2 \label{ourflatSchw}
\end{equation}
which is exactly the metric in the eq.(\ref{OURFREELYPADDYH}).
\end{itemize}

However, despite the difference between eq.(\ref{PADDYFLAT}) and eq.(\ref{ourflatSchw}), we will show that the latter 
exactly reproduces the results of the former, especially for the null vectors defined on $\mathcal{H}$ \cite{PaddyThermo}.\\
Given the metric in eq.(\ref{OURFREELYPADDY}) we can compute the vector normal to the worldlines of the comoving observer in the 
freely falling frame (so that 
we can well-define the horizon limit of this vector).\\
To define the normal vector $n^a$, we start defining the 
4-acceleration, $a^j$. Taking into account the coordinates in eq.(\ref{OURCOORDTRANSF}) we can differentiate them and it can be shown that 
the components of $a^j$ transform as the differentials, $dx$ and $dt$, 
for general transformations of coordinates, 
hence leading to a 4-acceleration with the following expression
\begin{equation}
 {a}\equiv {n}=\frac{(16m^2)^2}{(X^2-T^2)(X^2-T^2+16m^2)}(T\partial_T+X\partial_X)
\end{equation}
which, in this frame, exactly coincides with the normal vector $n$. Moreover 
its norm is $g({n},{n})=1$, viz., it is a spacelike vector.
We note that the denominator of this vector diverges when evaluated on the horizon, $X=T$, so $a$ and $n$ are ill-defined quantities 
on $\mathcal{H}$. For this reason we can define another vector
\begin{equation}
 {\xi} := \frac{X^2-T^2}{4m}a=\frac{(16m^2)^2}{(X^2-T^2+16m^2)^2}\frac{1}{4m}(T\partial_T+X\partial_X)\label{ourxi} 
\end{equation}
which limit on the horizon ($X=\epsilon T$) is well-defined and finite
\begin{equation}
 {\xi}\big|_{\mathcal{H}}=\frac{X}{4m}(\epsilon\partial_T+\partial_X)\label{ourxionH}
\end{equation}
where in the region we are interested, $|X|<T$, we take $\epsilon = \pm 1$. 
Its norm is also well-defined and finite on $\mathcal{H}$: so, when $T=\pm X$, using the eq.(\ref{ourflatSchw}), we get
\begin{equation}
\begin{split}
 g({\xi},{\xi})\big|_{\mathcal{H}}&=g_{TT}\epsilon^2X^2+g_{XX}X^2+2\epsilon X^2g_{XT}\\
 &=-\left(1-\frac{X^2}{8m^2}\right)X^2+\left(1+\frac{X^2}{8m^2}\right)X^2-\frac{\epsilon^2X^4}{4m^2}=0.\label{normourxionH}
\end{split}
 \end{equation}
So ${\xi}$ is a null vector on the horizon, viz., it is tangent to the horizon (and also normal, since $\mathcal{H}$ is a null hypersurface) 
and is well defined everywhere in the spacetime. It is exactly the null vector in eq.(\ref{Nn}) defined by Padmanabhan and in what follows we 
refer to it calling it ${n}$ (since ${n}$ and ${\xi}$ are related), for convenience.\smallskip

Finally, we can also show how vectors transform when we modify the metric, varying them, 
especially the null vectors normal to the black hole 
horizon, $n^a$. In fact, infinitesimal variations of the metric mean infinitesimal deformations of the parameters which define it, 
and we have seen that the deformation of $n^a$ plays a crucial role in the definition of the Padmanabhan's variational principle \cite{PaddyThermo},
for example when he 
adds a Lagrange multiplier to the entropy functional to impose that the null vector remains null after the variation. To see if this condition makes sense, 
we should show that the vector normal to the new horizon (which arises as a consequence of the deformation of the metric) 
is still a null vector on this surface.\\
In fact, let us consider a black hole described by a Schwarzschild metric defined by the parameter $m$, which is the mass of the black hole. 
When some amount of matter falls into the black hole, its mass grows up, and the metric describing the new black hole will be a Schwarzschild 
metric whose mass, $m$, will be related to the old one by the relation $m=\bar{m}(1+\delta m)$. Thus, since we are interested at what 
happens in the horizon, we must write the new Schwarzschild metric in the freely falling frame. Given the coordinate transformations 
in eq.(\ref{OURCOORDTRANSF}), these become
\begin{equation}
\begin{split}
 &T=\bar{T}(1+\delta m)=\bar{T}+\delta T\\
 &X=\bar{X}(1+\delta m)=\bar{X}+\delta X\label{NEWTX}
 \end{split}
 \end{equation}
where
\begin{equation}
\delta T = T\delta m, \qquad \delta X=X\delta m 
\end{equation}
that is the new coordinates are a function of the old one. With this coordinates, the difference between the new metric and the old one will be
\begin{equation}
\begin{split}
 g-\bar{g}=\delta g = &+\delta m\left[\frac{(16m^2)^2X^2+T^2(X^2-T^2+16m^2)^2}{(X^2-T^2)^2(X^2-T^2+16m^2)}\right]dT^2\\
 &-2XT\delta m\left[\frac{(16m^2)^2+(X^2-T^2+16m^2)^2}{(X^2-T^2)^2(X^2-T^2+16m^2)}\right]dXdT\\
 &+\delta m\left[\frac{(16m^2)^2T^2+X^2(X^2-T^2+16m^2)^2}{(X^2-T^2)^2(X^2-T^2+16m^2)}\right]dX^2. \label{DELTAGXT}
\end{split}
 \end{equation}
In this frame, before varying the mass, the null vector field $\vec{n}$ normal to the horizon, $\mathcal{H}$, was
\begin{equation}
 {n}|_{\mathcal{H}}=\frac{X}{4m}(\partial_T+\partial_X).
\end{equation}
Now, an infinitesimal transformation of coordinates of the type in eq.(\ref{NEWTX}) leads us to define the normal vector on the new 
horizon, $\mathcal{H}'$, and it corresponds to parallel transport this vector along a curve from a point in the old horizon to a point 
in the new one. Thus we need the expression of the Riemann curvature tensor, since 
the definition of the parallel transport around a loop depends on this tensor. We note that 
the new metric defers from the old one by an infinitesimal parameter, $\delta m$, so each tensor we can construct starting from $g'$ 
will depend on $\delta m$. The Riemann curvature tensor is of the second order in the derivatives of the metric, 
thus the
difference between the new curvature tensor and the old one will be at least of the second order in the derivatives of $\delta m$. 
It follows that, at the first order, this difference is negligible and this means that \textit{locally} the spacetime between $g$ and $g'$ is
flat, so the base vectors do not change when we go from $g$ to $g'$. For this reason, given ${n}_{\mathcal{H}}$ 
and ${n}_{\mathcal{H}}'$,
only comparing their components, we will expect the latter 
to have the same form of the former, viz., the same base vectors, since we aspect they do not change during this operation
(the difference between $T',X'$ and $T,X$ is a constant, so, at first order in $\delta m$, switching from the old frame to the new one 
the Jacobian of the difference vanishes). The vector field ${n}'$ normal to the new horizon $X^2=T^2+16m^2\delta m^2$ 
will get the following components:
\begin{equation}
 {n}'|_{\mathcal{H}}=(1+\delta m)\frac{X}{4m}(\partial_X+\partial_T)= (1+\delta m){n}|_{\mathcal{H}}.
\end{equation}
Thus, to change the mass in the Schwarzschild metric means to 
modify the length of this vector, dilating it. 
Moreover, by construction, ${n}_{\mathcal{H}}'$ is still a null vector on the new horizon, 
since ${n}_{\mathcal{H}}$ is a null vector field on
$\mathcal{H}$ and their difference, ${n}_{\mathcal{H}}'-{n}_{\mathcal{H}} = \delta m {n}_{\mathcal{H}}$, 
depends on a constant parameter which - in turn - does not modify the nature of this vector field on $\mathcal{H}'$.\\
For example, this geometrical result also explains why Padmanabhan can add a Lagrangian multiplier to its entropy functional \cite{PaddyThermo} 
to take into account the condition that the normal vector field $n^a$ still remains a null vector after its variation.




\section*{Acknowledgments}
This article is based upon work from COST Action (CA15117 CANTATA), supported by COST (European Cooperation in Science and Technology). 
We acknowledge Professor M. Cadoni for useful discussions about these topics and the 
contribution of INFN (Iniziativa Specifica QGSKY), 
the local research project {\it Metodi Geometrici in Fisica Matematica e Applicazioni} (2015) 
of Dipartimento di Matematica of University of Torino (Italy).

\end{document}